\documentclass[12pt,preprint]{aastex}
\usepackage{graphics}
\input epsf
\begin{document}

\title{A Turbulent Origin for Flocculent Spiral Structure in Galaxies}

\author{Bruce G. Elmegreen \affil{IBM Research Division, T.J. Watson
Research Center, P.O. Box 218, Yorktown Heights, NY 10598, USA,
bge@watson.ibm.com} }

\author{Debra Meloy Elmegreen \affil{Vassar College,
Dept. of Physics \& Astronomy, Poughkeepsie, NY 12604;
elmegreen@vassar.edu} }

\author{Samuel N. Leitner \affil{Wesleyan University, Dept. of Physics, Middletown,
CT; sleitner@wesleyan.edu}}

\begin{abstract} The flocculent structure of star formation in
galaxies has a Fourier transform power spectrum for azimuthal
intensity scans with a power law slope that increases systematically
from $\sim -1$ at large scales to
$\sim -5/3$ at small scales. This is the
same pattern as in the power spectra for azimuthal scans of HI
emission in the Large Magellanic Clouds and for flocculent dust clouds
in galactic nuclei. The steep part also corresponds to the slope of $\sim-3$
for two-dimensional power spectra that have been observed in
atomic and molecular gas surveys of the Milky Way and the Large
and Small Magellanic Clouds.  The power law structure for star
formation in galaxies arises in both flocculent and grand design
disks, which implies that star formation is the same in each and
most likely related to turbulence. The characteristic scale
that separates these two slopes
corresponds to several tens of pixels or
several hundred parsecs in most galaxies, which is comparable to
the scale height, the inverse of the Jeans wavenumber and the
size of the largest star complexes. We suggest that the power
spectrum of optical light is the result of turbulence,
and that the large-scale turbulent motions
are generated by sheared gravitational instabilities which
make flocculent spiral arms first and then
cascade to form clouds and clusters on smaller scales.
Stellar energy sources presumably
contribute to this turbulence by driving smaller scale
motions and by replacing the gravitational
binding energy that is released during spiral arm collapse. The
spiral wavemode in the image of M81 is removed by reconstructing
the Fourier transforms without the lowest 10 wavenumbers.  The
result shows the underlying flocculent spirals and reverse-shear
spirals of star formation that are normally overwhelmed by the
density wave.
\end{abstract}
\keywords{turbulence --- stars: formation --- ISM: structure --- galaxies: star clusters --- galaxies: spiral}

%\keywords{turbulence --- stars: formation ---
%ISM: structure --- galaxies: star clusters}

\section{Introduction}

Spiral structure in galaxies is a combination of sheared features
from star formation and spiral density waves in the gas and stars.
Most non-barred galaxies in the field, such as NGC 2841 and NGC
5055, have short and patchy arms (``flocculent''), while most
galaxies with bars or nearby companions, such as M81 and M51, have
spiral waves or wavemodes that give them a grand design (Kormendy
\& Norman 1979; Elmegreen \& Elmegreen 1982; Elmegreen, Elmegreen
\& Seiden 1989). Flocculent galaxies also have small disk-to-halo
mass ratios (Athanassoula, Bosma, \& Papaioannou 1987; Elmegreen
\& Elmegreen 1990; Fuchs 2002), which helps to explain why their
spiral waves are weak (Mark 1976; Thornley 1996; Kuno, et al.
1997; Elmegreen et al. 1999; Puerari et al. 2000).

Star formation in grand design galaxies is often attributed to
compression from a density wave (Roberts 1969), although these
galaxies have no higher star formation
rates per unit gas or area than flocculent
galaxies (Elmegreen \& Elmegreen 1986; Stark, Elmegreen, \& Chance
1987). Both types also satisfy the Kennicutt (1998) relation
between average star formation rate and gas surface density. These
observations, along with the similar star formation properties in
dwarf irregular galaxies (Hunter \& Gallagher 1986) and the
decoupling between B and K-band features in galaxy disks (Block \&
Puerari 1999) suggest that small-scale star formation processes
are the same everywhere, regardless of the presence or lack of a
spiral wave.  This means that if a strong wave is present and a
high fraction of the star-forming clouds are formed by the wave,
then the star formation processes inside these clouds will be no
different than the star formation processes elsewhere. It also
means that if there is no stellar wave, then the interstellar gas
will form the same total surface density of clouds by other means
(see review in Elmegreen 2002).

The nature of the star formation process itself has been
elucidated recently through studies of geometry and
kinematics. Fractal-fitting procedures based on star-neighbor
distributions, cell counting, and cluster hierarchies (Feitzinger
\& Braunsfurth 1984; Feitzinger \& Galinski 1987; Battinelli,
Efremov, \& Magnier 1996; Elmegreen \& Salzer 1999;
Heydari-Malayeri et al. 2001; Pietrzynski et al. 2001), unsharp
masks (Elmegreen \& Elmegreen 2001), and auto-correlation
functions (Feitzinger \& Braunsfurth 1984; Harris \& Zaritsky
1999; Zhang, Fall, \& Whitmore 2001) show a hierarchical,
self-similar type of structure that often extends from the
telescope resolution up to a kiloparsec or more. Self-similar
structure also goes to sub-parsec scales in nearby regions (Testi
et al. 2000). These patterns are apparently present as long as the
regions are younger than a crossing time (Elmegreen 2000); after
this, random stellar motions smooth out the geometric structure of
stellar birth.

The hierarchical structure for star formation is similar to that
in the interstellar gas, which is also fractal over a wide range
of scales (Dickman, Horvath \& Margulis 1990; Scalo 1990;
Falgarone, Phillips, \& Walker 1991; Elmegreen \& Falgarone 1996;
St\"utzki et al. 1998; Westpfahl et al. 1999; Keel \& White 2001).
This similarity suggests that young stars follow the gas as they
form; some are triggered in this gas by external stellar pressures
and others form spontaneously. There is also a time and distance
correlation for young stars that is similar to that for the gas
(Efremov \& Elmegreen 1998). This implies that stars initially
move along with the gas and that both the star formation event and
the cloud lifetime are limited by the transience of local gas
flows (Ballesteros-Paredes, Hartmann, \& V\'azquez-Semadeni
1999).

The origin of the kinematic and spatial correlations is probably
turbulence (Larson 1981; Falgarone \& Phillips 1990). Numerical
simulations reproduce these features well and also show how stars
might form spontaneously in a turbulent medium (Klessen, Heitsch,
\& Mac Low 2000; Ossenkopf, Klessen, \& Heitsch 2001; Klessen
2001; Padoan et al. 2001a; Padoan \& Nordlund 2002).

Interstellar turbulence can arise from a combination of
spiral-forming instabilities driven by the swing-amplifier (Toomre
\& Kalnajs 1991; Thomasson, Donner, \& Elmegreen 1991; de Vega,
Sanchez, \& Combes 1996; Fuchs \& von Linden 1998; Semelin \&
Combes 2000; Crosthwaite, Turner, \& Ho 2000; Bertin \& Lodato
2001; Wada \& Norman 2001; Wada, Meurer, \& Norman
2002; Huber \& Pfenniger 2001a,b, 2002;
Vollmer \& Beckert 2002; Chavanis 2002) and stellar pressures
(Norman \& Ferrara 1996; Avila-Reese \& V\'azquez-Semadeni 2001).
Spiral instabilities in the gas pump gravitational binding energy
and shear energy into turbulent motions over a wide range of
scales. Stellar pressures may generate turbulence only locally
(Braun 1999; V\'azquez-Semadeni \& Garc\'ia 2001). Magnetic
instabilities can also generate turbulence (Asseo et al. 1978;
Sellwood \& Balbus 1999), but they generally do not produce dense
cloudy structures by themselves.

If spiral-forming instabilities drive the turbulence and
compression that make star-forming clouds, and if this turbulence
also produces correlated structures that eventually show up in
young star fields, then flocculent spiral structure in galaxies
should have the same geometric properties as local interstellar
clouds.  If star formation processes are also independent of
stellar spiral waves, then the star-forming parts of a grand
design galaxy should have the same intrinsic geometry as
flocculent spirals.

This paper measures the Fourier transform power spectra of optical
light in six galaxies and compares them with the power spectra of HI
emission from the LMC.  The spectra are very similar, suggesting that
star formation follows a turbulent gas.  Optical light is more complex
than HI emission, however, because dust and bright stars dominate the
high spatial frequencies, stars have a mixture of ages, and there is no
velocity information.  Nevertheless, optical passbands have much higher
spatial resolution than radio, CCD images are available for most galaxies,
and most of the features they show are closer to the actual star formation
process than large-scale gas surveys, which have revealed primarily the
low or moderate density phases so far.  Thus the power spectrum of optical
light might reveal a connection between star complexes and spiral arms,
but it probably cannot distinguish between different models of turbulence,
which all have similar power spectra.

Optical power spectra should also give a physical scale at which star
formation becomes coherent. This scale could be several hundred parsecs
to a kiloparsec, corresponding to the disk thickness and inverse Jeans
wavenumber (Zhang, Fall, \& Whitmore 2001).  If this is the driving scale
for the turbulence that generates star-formation structures, then the
source of this turbulence would be the spiral instabilities that make
flocculent arms. Turbulence driving by spiral instabilities could also
occur at smaller scales, followed by an inverse cascade in two dimensions
(Wada, Meurer, \& Norman 2002), but our observations of features on
small scales are hampered by pixel noise and a possible transition from
2-D to 3-D turbulence in the real interstellar media of these galaxies.

In the next section, we review the suggested connections between
turbulent motions and power spectra of density or column density.
Section \ref{sect:lmc} shows the normalized power spectrum of HI
emission from the LMC, multiplied by $k^{5/3}$ to flatten the
power-law part for easier interpretation. Then Section
\ref{sect:data} describes our galaxy data, section \ref{sect:six}
shows power spectra of optical emission from these galaxies, and
section \ref{sect:energy} discusses the likely energy sources for
the turbulence that gives star formation its structure.  In
Section \ref{sect:m81}, we remove the lowest Fourier components
from the grand design galaxy, M81, and show that the star
formation patterns resemble those in a flocculent galaxy,
aside from the obvious orbit crowding and reverse shear that
accompanies the density wave. The results are summarized in
Section \ref{sect:sum}.

A second paper (Elmegreen, et al. 2003) determines the optical power
spectra for another galaxy, M33, which has a larger dynamic range
for power law analysis than the current sample because of its smaller
distance.  This second paper also models the result in an attempt to
explain the entire spectrum, rather than just the $k^{-5/3}$ part that
is emphasized here.

\section{Density Power Spectra and Turbulence}
\label{sect:density}

One way to observe the correlated structure in a turbulent gas is
with a Fourier transform power spectrum, which is the sum of the
squares of the real and imaginary parts of the Fourier transform
of the emission, plotted as a function of the spatial frequency,
$k$. For two or three dimensions, the spatial frequency is the
quadratic sum of the spatial frequencies in each direction.
The two-dimensional power spectra of Milky Way HI emission
(Crovisier \& Dickey 1983; Green 1993; Dickey et al. 2001), HI
absorption (Deshpande, Dwarakanath, \& Goss 2000), CO
(St\"utzki et al. 1998), and IRAS 100 micron emission (Gautier, et
al. 1992; Schlegel, Finkbeiner, \& Davis 1998) are power laws with
slopes of $-2.8$ to $-3$. The same power laws were found for
two-dimensional HI emission from the Small and Large Magellanic
Clouds (Stanimirovic, et al. 1999; Elmegreen, Kim, \&
Staveley-Smith 2001) and for dust emission from the Small
Magellanic Cloud (Stanimirovic et al. 2000).

Power spectrum analyses of data cubes require special care to
account for sharp boundaries.  For example, Dickey et al.
(2001) multiplied the emission at the edge of their survey by a
Gaussian smoothing function.  One way to avoid this problem is
with the Delta Variance technique (St\"utzki et al. 1998;
Zielinsky \& St\"utzki 1999; Ossenkopf, Klessen, \& Heitsch 2001;
Bensch, St\"utzki, \& Ossenkopf 2001), which does not try to fit a
periodic function to the data.

For galaxies, the two-dimensional Fourier transform has an
additional problem in that it includes the structure from the
exponential disk, which is generally not wanted in an analysis of
interstellar clouds.  For the Small Magellanic Cloud studied by
Stanimirovic et al. (1999), this was not an issue because there is
no disk. The LMC does have this problem, though, and for this
reason, Elmegreen, Kim, \& Staveley-Smith (2001) used
one-dimensional Fourier transforms of the azimuthal profiles.
Azimuthal profiles are ideal for Fourier transforms because they
are periodic and the systematic radial gradients in the disk do
not contribute. One-dimensional power spectra have shallower
slopes than two-dimensional spectra by one unit, so the azimuthal
power spectrum of gas emission from a galaxy becomes a power law
with a slope of $\sim-2$.

The connection between these power spectra for intensity or column
density and the power spectrum of velocity fluctuations associated
with turbulence is not obvious. Lazarian \& Pogosyan (2000) showed
that the structure of line emission from a turbulent gas is the
result of a combination of density and velocity irregularities,
with the contribution from each depending on the ratio of the
integrated width over the spectrum to the turbulent line width for
the scale observed.  They considered the power spectrum for
three-dimensional density alone to be $P_{den,3}(k) \propto
k^{-n}$, and the power spectrum for three-dimensional velocity
alone to be $P_{vel,3}\propto k^{-3-m}$ for wave number
$k=2\pi/\lambda$ and wavelength $\lambda$.  In this notation,
Kolmogorov turbulence would have $m=2/3$.  In the case of a
shallow density spectrum, where $n<3$, the power spectrum of the
{\it projected} two-dimensional line emission varies from
$P_{em,2}\propto k^{-n+m/2}$ for a thin slice of the velocity
spectrum (i.e., a single velocity channel width) to a steeper
function, $\propto k^{-n}$, for a slice that is wider than the
velocity dispersion on the scale $1/k$.  In the case of a steep
density spectrum, $n>3$, the 2-D projected power spectrum varies
from a different thin-slice limit, $P_{em,2}\propto k^{-3+m/2}$,
to a steeper thick-slice function, $\propto k^{-3-m/2}$, to the
same very-thick slice limit, $\propto k^{-n}$. In both cases, the
thick-slice limit has the same projected two-dimensional power
spectrum as the pure density spectrum in three dimensions.  Also
in both cases, the shallowness of the thin-slice limit implies
there is a lot of small scale projected structure in velocity
channels that is purely the result of velocity crowding on the
line of sight, i.e., structure that has no density counterpart and
therefore is not from physical objects like ``clouds'' in the
usual sense (see also Ballesteros-Paredes, V\'azquez-Semadeni, \&
Scalo 1999; Pichardo, et al. 2000).

If there are no velocity fluctuations, then the two-dimensional
power spectrum of projected emission from a physically thin slab
on the line of sight (thinner than $1/k$) is $P_{den,2,slab}
\propto k^{1-n}$, which is shallower than the 3-D density spectrum
by one unit (Lazarian et al. 2001).

Stanimirovic \& Lazarian (2001) found this predicted steepness
variation with slice thickness in the HI emission-line power
spectra from the Small Magellanic Cloud, giving $n=3.3$ for the
pure density index and $-3-m=-3.4$ for the pure velocity index.
This latter index is slightly different than the power spectrum
index for pure velocity in incompressible Kolmogorov turbulence,
which is $-11/3$.  Lazarian et al. (2001) also did numerical
simulations of 3-D compressible turbulence in a self-gravitating,
non-magnetic gas with star formation and Coriolis forces
(representing a 300 pc piece of a galaxy) and got the predicted
power spectrum for thin slices and the predicted slope variation
with velocity-slice thickness.

Goldman (2000) considered only the velocity integrated HI
structure (thick-slice limit) in the SMC observations by
Stanimirovic et al. (2000) and assumed the projected density
structure (``clouds'') is the result of a passive response to 3-D
incompressible turbulent motions (see also Higdon 1984; D\'ecamp
\& Le Bourlot 2002). This would be analogous to the case of a dye
tracer in an incompressible turbulent fluid. In this case, the
two-point correlation function of the density is proportional to
the two-point correlation function of the velocity, and the
density power spectrum has the same slope as the velocity power
spectrum. Goldman showed that the HI emission always comes from
cloudy structures that are very thin compared to the line of sight
through the whole galaxy and because of this the power spectrum of
the projected structure, with its observed slope of $\sim-3$,
corresponds to a power spectrum of 3-D velocity structure that has
a slope of $-4$ (in the notation above). This is slightly steeper
than the 3-D velocity structure in Kolmogorov turbulence, which
would have a slope of $-11/3=3.67$, whereas the 3-D velocity
spectrum derived by Stanimirovic \& Lazarian (2001) for the SMC,
with its slope of $-3.4$, is slightly shallower than Kolmogorov.
Goldman (2000) interpreted his steep result as an effect of energy
dissipation during the cascade of energy from large to small
scales, and noted that observations by Roy \& Joncas (1985) and
simulations by Vazquez-Semadeni, Ballesteros-Paredes, \& Rodriguez
(1997) got this result too.  A medium consisting entirely of shock
fronts would have a velocity power spectrum of the same slope
(V\'azquez-Semadeni, et al. 2000).

Goldman (2002) also found that the power spectrum of density in
his model steepens by one unit when the inverse wavenumber scale
becomes smaller than the cloud-layer thickness, $L$.  This implies
that the two-dimensional power spectrum of density for a passive
scalar in a Kolmogorov turbulent fluid should have a slope of
$-8/3$ for transverse scales larger than the thickness, where
$k<1/L$, and a slope of $-11/3$ for transverse scales smaller than
the thickness, where $k>1/L$.  One-dimensional strips through such
a medium would therefore have power spectrum slopes that steepen
from $-5/3$ on large scales to $-8/3$ on small scales, i.e.,
differing by 1 in both cases. Goldman (2002) did not consider that
this steepening would be observed in galaxies, but it may have
been observed already in the LMC HI emission (Elmegreen et al.
2001) and we attempt to observe it here again for optical galactic
emission.  If it can be observed, and if Goldman's theory applies,
then we have a good method for obtaining the line-of-sight
thickness of a galaxy based entirely on the transverse patterns of
optical starlight emission.

Analysis techniques other than Fourier transform power spectra
have also been used to relate velocity and density structures in a
turbulent medium.  For example, the structure function of an
extinction map of the Taurus region was found to compare well with
the structure function for velocity in supersonic turbulence
(Padoan, Cambr\'esy \& Langer 2002).  This function involves the
average value of the difference between the extinction or the
velocity, respectively, at all pairs of points separated by some
distance and raised to a power.  The density and velocity
correspondence in this case was not explained, however, but it
could be another manifestation of density as a passive tracer for
a type of turbulence that acts in a nearly incompressible way. For
example, Boldyrev, Nordlund, \& Padoan (2002) propose a model for
explaining structure functions in which turbulence is mostly
solenoidal in the inertial (power-law) range and then becomes
Burgers-like (sharp-edged and shock-dominated) on small scales
where dissipation occurs; such a model makes density a passive
scalar in the inertial range. Another measure of structure is the
probability distribution function for column density, which spans
a range between log-normal and Gaussian depending on the number of
independent correlated regions on the line of sight
(V\'azquez-Semadeni \& Garc\'ia 2001).  Analysis techniques based
on spectral line information, such as principal component
analyses (Brunt \& Heyer 2002) or the spectral correlation function
(Padoan, Rosolowsky, \& Goodman 2001)
are not useful for optical images.

Evidently, the physical processes that give rise to the density
structure of interstellar gas are not completely determined. A
medium filled with numerous cool clouds that are moved in bulk by
gravitational forces associated with swing-amplified instabilities
and by pressure forces associated with star formation should have
some of the properties of a passive scalar convected by
independent turbulent motions. If this turbulence is somewhat
incompressible in the inertial range, then the two-dimensional
power spectrum for projected column density can have a slope of
$-11/3$ to $-4$, depending on the importance of energy losses in
the cascade to small scales, and on the prevalence of shock fronts
and other sharp edges from ionization and stellar wind
interactions.  These slopes increase by 1 for one-dimensional
scans, and they can increase by 1 again if the transverse
dimension measured by $1/k$ is much larger than the line-of-sight
thickness.

\section{Power Spectrum of HI in the Large Magellanic Cloud}
\label{sect:lmc}

The power spectrum of HI emission from azimuthal scans in the LMC
is shown on the left in Figure \ref{fig:lmc}, using data from
Elmegreen et al. (2001). The curves are shifted upward for
increasing galactocentric radii, as indicated. The wavenumber on
the abscissa is normalized to the maximum value, which is the same
for each curve and corresponds to an image scale of 2 pixels. The
power spectra for large radii cover a wider wavenumber range than
the power spectra for small radii because the scan length is
larger at larger radii so the minimum wavenumber, normalized to
$1/\left(2\;{\rm px}\right)$, decreases inversely with radius. A
similar diagram was shown in Elmegreen, Kim, \& Staveley-Smith
(2001) but with an error at the highest frequency inadvertently
caused by interpolation between pixels. The azimuthal scan of
noise from beyond the galaxy edge produces a power spectrum with a
slope of $-1$, as shown by the dashed line. The slope at
intermediate $k$ in each scan is $\sim-5/3$ and the slope at high
$k$ up to $k\sim0.5$ is $\sim-8/3$, as indicated by the straight
lines at the top of the figure. The highest wavenumbers have a
slope of about $-1$ again, and therefore look dominated by noise
(this flattening at high $k$ was also found by Lazarian et al.
2001 in numerical simulations). The transition from $-5/3$ to
$-8/3$ at physical scales of $\sim100$ pc was interpreted by
Elmegreen et al. (2001) to be an indication of the line-of-sight
disk thickness, following predictions by Lazarian \& Pogosyan
(2000) and Goldman (2000). Padoan et al. (2001b) found a similar
transition for this LMC data using a different technique.

The variations in slope for these power spectra are difficult to
see by eye, so we multiply each by $k^{5/3}$, which has the effect
of flattening the inertial range to a horizontal line. The
right-hand side of Figure \ref{fig:lmc} shows these normalized
power spectra. The short lines near some of the spectra at high
frequency have a slope of $-1$ in the normalized plot, which
corresponds to a slope of $-8/3$ in the power spectrum; these
lines indicate what the $-5/3$ to $-8/3$ transition should look
like in a normalized, one-dimensional power spectrum.

\section{Galaxy Data}
\label{sect:data}

The galaxies in Table 1 were chosen for their diverse spiral types
and because of the large number of pixels in the available images,
shown in Figure \ref{fig:images}. Most of the data were obtained
from the Hubble Space Telescope's (HST) online archive
(www.stsci.edu), with the exception of NGC 3031 (M81), which was
obtained by J.C. Cuillandre in 1999 at the Canada-France-Hawaii
Telescope (Cuillandre, et al. 2001). The HST exposures were first
combined to reduce gamma rays and improve signal-to-noise, and
then they were mosaicked in IRAF. The complete images were
transformed through {\it FV viewer}\footnote{This research made
use of the software package {\it FV Viewer}, obtained from the
High Energy Astrophysics Science Archive Research Center
(HEASARC), provided by NASA's Goddard Space Flight Center.} into
ASCII intensity tables and processed with a Fortran program. In
the cases of M81 and NGC 7742, star reduction algorithms were used
to eliminate some foreground stars.

The Fortran program took azimuthal intensity scans in a
deprojected image at equally spaced radii around the center of
each galaxy. The radial intervals ($\Delta R$) are given in Table 1.
Inclination angles were taken from the Third
Reference Catalogue of Bright Galaxies (de Vaucouleurs et al.
1991; hereafter RC3); position angles were calculated from the V3
angle of HST, as given in the FITS header, and from the position
angle from north, as given by the RC3.

The azimuthal scans were sampled at one pixel spacing with no
interpolation between pixels.  Sample intensity scans at the
radius $R=2.03$ kpc in M81 are on the bottom in Figure
\ref{fig:azim}. Each curve is an average of 9 azimuthal scans
separated in radius by one pixel. The original one on the left has
a spike from a bright star at an azimuthal position of 926 px; the
cleaned scan on the right has the star partially removed. Most
galaxy images did not have bright foreground stars like M81, so
they were not cleaned.

Fourier cosine and sine transforms, ${\hat I}_c(k)$ and ${\hat
I}_s(k)$, were taken of each intensity scan, $I(\theta)$, using
the direct sums,
\begin{equation}
{\hat I}_c(k)=\sum_{n=1}^N \cos(k 2\pi n/N)I(n)\end{equation}
\begin{equation}
{\hat I}_s(k)=\sum_{n=1}^N \sin(k 2\pi n/N)I(n).\end{equation}
Here, $N$ is the number of pixels in the azimuthal scan and $k$ is
the wavenumber. An FFT algorithm was not used because that would
overly constrain the scan lengths. The power spectrum is the sum
of the squares of these transforms:
\begin{equation}P(k)={\hat I}_c(k)^2+{\hat I}_s(k)^2\end{equation}
To reduce noise, the power spectra from 9 adjacent azimuthal scans
separated by one pixel in radius were averaged together for all
the results here.

Figure \ref{fig:ellipse} shows the azimuthal scans in M81 and
NGC 5055 as semi-transparent ellipses. Each ellipse consists of
the 9 adjacent azimuthal scans that were used for the average
power spectra.

The average power spectrum at the radius of the average azimuthal
scan on the left in Figure \ref{fig:azim} is shown in the panel
above it as the top curve. The power spectrum is highly distorted
because of the star. The curve below it is the power spectrum
multiplied by $k^{5/3}$. This normalization will be done for all
of our power spectra in order to make it easier to recognize
wavenumber regions where there is something like Kolmogorov
turbulence.

The top curve on the right of Figure \ref{fig:azim} shows the
power spectrum of the cleaned scan. It is nearly a smooth power
law with a slope of $-1.45$.  The curve below it is again
normalized by the multiplicative factor of $k^{5/3}$.

\section{Power Spectra for Six Galaxies}
\label{sect:six}

The power spectra for 13 equally spaced radii in M81 are shown in
the left-hand panel of Figure \ref{fig:psm81}, stacked up with
increasing radius toward the top.  The normalized, star-subtracted
curve from the right-hand side of Figure \ref{fig:azim} is the 4th
scan up from the bottom in Figure \ref{fig:psm81}.  The other
radii in M81 were corrected for obvious stars too.  A star was
defined algorithmically to be a pixel more than 1.2 times brighter
than the average of the pixels around it. There is still some
distortion in the normalized power spectra in Figure
\ref{fig:psm81} because of unresolved bright sources, most of
which are inside M81. Generally the normalized power spectra are
approximately flat between 20 pc and 100 pc. This means the original power
spectra have slopes near $-5/3$ in this wavenumber region, making
them resemble density structures from
Kolmogorov turbulence, or within the uncertainties,
Burgers turbulence (giving a slope of $-2$).

At large radii, the normalized power spectra have slopes near
$\sim2/3$, as indicated by the dashed line, which means the
original power spectra have a slope of $\sim-1$. Because this
slope occurs in the outer parts of galaxies where the intensity is
weak, it is probably from a combination of noise, random positions
for remote flocculent arms, and weak long-range correlations.
Models in Elmegreen et al. (2003) fit both the $-1$ and $-5/3$
slopes for globally correlated structures with a single intrinsic
power spectrum slope of $-5/3$. The relatively shallow part on
large scales in the models is partly the result of using azimuthal
scans rather than linear scans, i.e, flocculent arms are poorly
correlated on opposite sides of the galaxy and well correlated
with their immediate neighbors on the same side. The models also
show a resolution effect in which the extended $-5/3$ part comes
partly from unresolved structures like clusters and OB
associations, which have a range of sizes around the 100 pc scale
we measure here.

Power spectra at 10 equally spaced radii for the other galaxies
are shown in Figures \ref{fig:psn5055} to \ref{fig:psn7742}. NGC
5055 and NGC 7742 have data for two passbands.  The left-hand
panel of Figure \ref{fig:psn5055} has a 3-segment line to guide
the eye. The rising part of the left segment has a slope of $2/3$,
representing noise, the flat part in the center has a slope of 0,
representing the projected density in the
Kolmogorov part of a power spectrum for 2-D
turbulence, and the falling part on the right has a slope of $-1$,
representing the projected density in the
Kolmogorov part of a power spectrum for 3-D
turbulence.

The results of the power spectrum analysis are summarized in Table
2, which gives the size ranges for the horizontal parts of the
normalized power spectra in parsecs and in proportion to R$_{25}$,
as indicated by the scales at the tops of the figures.

There is little difference in the power spectra for flocculent and
grand design galaxies.  NGC 5055 is a flocculent galaxy and its
power spectrum in Figure \ref{fig:psn5055} is similar to that of
M81 in Figure \ref{fig:psm81} except for the strong components in
M81 at low $k$, which are from the spiral density wave.

The maximum size of the horizontal part of the normalized power
spectrum is generally larger at smaller radii than larger radii,
where noise seems to dominate small $k$.  Most galaxies have
unresolved bright features that are intrinsic to the galaxy, in
addition to faint foreground stars, and these distort the power
spectra at high $k$ (as in Figure \ref{fig:azim}). Between these
wavenumber limits, four galaxies have clear regions where the
normalized power spectra are approximately
horizontal in these figures: M81, NGC
4414, NGC 5055, and NGC 7217. These horizontal parts suggest that
the non-normalized power spectra decrease as $k^{-5/3}$, which is
appropriate for column density structure in
Kolmogorov turbulence in a one-dimensional scan.
Unfortunately, the uncertainties from brightness variations among stars,
pixel noise, and the limited power law range, make it
impossible to distinguish between this slope and
$-2$, for example, so we cannot specify what type of turbulence
might be causing these brightness patterns.

The other galaxies could have the same structure but be too noisy
or distant to resolve here.  In only one case, NGC 5055, is the
pixel and intrinsic stellar noise low enough to see a possible
transition to a 3-D column density spectrum at high frequency.  This
galaxy differs from the others in having very bright patches of
well-resolved star formation throughout the disk. NGC 7742 has
faint patches everywhere but in a ring of radius $\sim800$ pc,
where they are bright, and there is no evidence for a $k^{-5/3}$
spectrum anywhere except possibly in this ring (the third
scan up from the bottom); unfortunately, this ring also has so
many bright pixels that the power spectrum is distorted.
Presumably the distance to NGC 7742 is too large to resolve the
several-hundred pc outer scale for turbulence, so each star
complex with this scale has a size of only a few tens of pixels.
NGC 7217 has the same problems as NGC 7742: it is too distant to
resolve the turbulent structures and the star-formation patches
are intrinsically faint.

The meaning of the length scale at the top of the figures is $1/k$
in the azimuthal direction. Because 1 on the lower abscissa
corresponds to a wavenumber of $1/\left(2\;{\rm px}\right)$, the
length scale at the top is given by $L=2\theta D$ divided by the
normalized scale at the bottom, for pixel size $\theta$ in radians
and galaxy distance $D$. This azimuthal length is not the same as
the total feature length, which, because of shear, can be much larger if it is
measured along a spiral arm.
The typical length scale in the flat part of our power spectra
is 100 pc, with a range between 50 pc and 1000 pc, depending on galaxy and
galactocentric radius.
This distance is comparable
to the inverse of the Jeans wavenumber in the interstellar gas,
which is $c^2/\left(\pi G\Sigma\right)\sim200$ pc for velocity
dispersion $c=5$ km s$^{-1}$ and disk mass surface density
$\Sigma=0.002$ g cm$^{-2}\sim10$ M$_\odot$ pc$^{-2}$. Star
formation regions much larger than this in the azimuthal direction are
apparently not coherent on average (although a few
individual regions might be), and those separated by more than this
distance, such as those on opposite sides of the galaxy,
are independent. Star-forming regions smaller than $\sim100-500$ pc
tend to be correlated.

The maximum scale for the $k^{-5/3}$ part is also comparable to the
perpendicular disk scale height.  Either this height, along with
the Jeans length, is a true maximum scale for star formation (as
suggested by Efremov 1995 and others), or correlated regions with
larger sizes are so much older, redder, and dimmer that they do
not stand out in our optical images.  A power spectrum analysis of
very sensitive K-band images might find longer range correlations.
According to the time-size correlation in Efremov \& Elmegreen
(1998), regions 1 kpc in size, such as Gould's Belt, form stars
for $\sim30$ My, and these are still visible in our optical
images. An extrapolation of this relation to regions 10 kpc in
size suggests a formation time over $\sim100$ My, in which case
they should be dominated by stars of type A. A-star groupings of
this size may exist without detection in the optical data used
here. They would be relatively smooth, dim, and sheared into tight
spirals.  On the other hand, it is plausible that correlated
regions much larger than a scale height do not exist because
turbulent motions parallel the plane, driven by spiral
instabilities for example, cannot be much larger than turbulent
motions perpendicular to the plane, which would have to be driven
by 3-D mixing of the in-plane motions.  A fixed ratio for these
two speeds was suggested in simulations of spiral instabilities by
Huber \& Pfenniger (2001a).

\section{Energy Source for Interstellar Turbulence}
\label{sect:energy}

If the gravitational Jeans length or the scale height in the
ambient medium is the typical scale for coherent, star-forming
structures, then the implication is that gaseous self-gravity
initiates the formation of star-forming clouds, producing star
complexes.  A turbulent cascade to smaller scales forms smaller
clouds and smaller stellar conglomerates, such as OB associations
and subgroups, while shear stretches the complexes on larger
scales into flocculent spirals. This is beginning to be the
standard model for star formation but now it has the additional
aspect that all resolvable scales inside the unstable wavelength
are hierarchically fractal for both the gas and the stars.

This origin of structure on the Jeans length also implies that
much of the interstellar turbulence is from gravitational
instabilities in the gas.  This is consistent with recent
numerical experiments (see references in the Introduction). The
turbulent energy comes directly from gravitational binding energy
of the gas layer. This is not a renewable energy source because
collapse dissipates energy by radiation, so ultimately the energy
of turbulence has to come from shear or stars. Shear is not
important on the scale of individual clouds because most of them
are denser than the gravitational tidal density
($=3A\Omega/\left[\pi G\right]$ for Oort $A$ and angular rate $\Omega$),
so they cannot be broken apart
by shear.  This is unlike the case for whole flocculent spirals,
which are easily stretched by shear. Numerical simulations with
only self-gravity and shear get renewable spirals for this reason
(Toomre \& Kalnajs 1991; Sellwood \& Carlberg 1984; Semelin \&
Combes 2000; Huber \& Pfenniger 2002), but analogous simulations
where dense clouds form cannot break these clouds apart without
something like star formation and young stellar pressures (Huber
\& Pfenniger 2001a; Wada \& Norman 2001). Viscosity from
cloud-cloud interactions in the presence of shear can also break
apart spiral instabilities (Vollmer \& Beckert 2002).

The energy liberated by continuous spiral instabilities is
important as a source of turbulence in the ISM (e.g., Vollmer \&
Beckert 2002; Wada, Meurer, \& Norman 2002). The power density is
approximately the ISM energy density $\rho c^2$ multiplied by the
instability growth rate, $\pi G\Sigma/c$, which amounts to
$2\times10^{-27}$ erg cm$^{-3}$ s$^{-1}$ in the Solar
neighborhood. This power input is important because the spiral
growth time is comparable to crossing time over the disk
thickness, and this is about the energy dissipation time from
shocks and turbulent decay. The energy is comparable to that from
supernovae at a rate of one per 100 years in a whole galaxy,
assuming a 1\% efficiency for converting the explosion energy into
ISM motions.  Both spiral instabilities and stars provide enough
power to interstellar turbulence and cloud formation, but {\it spiral
instabilities may be preferred because they are pervasive,
large-scale, and cold}, which allows the gravity-driven motions to
be supersonic and strongly compress the gas.  Stellar power is
hotter and more localized, producing ionized and hot uniform
cavities with cloudy debris at the edge in a turbulent shell or
layer. The primary role of stellar energy on a large scale is to
provide internal energy that reverses the cold collapse of cloud
and star formation, thereby maintaining a steady state in the
presence of continuous spiral instabilities.

Spiral instabilities can also move cold clouds around without much
feedback from them in the form of pressure gradients. This can
produce a type of structure that is analogous to a passive scalar
in laboratory Kolmogorov turbulence.  The sharp edges in these
cold clouds that are produced by ionization and stellar winds can
steepen a three-dimensional density power spectrum from $-11/3$ to
$-4$, as discussed in Section \ref{sect:density}, but such sharp
edges are not likely to show up in the stars unless very young
regions are observed, as in the compressed ridge of Orion
(Reipurth, Rodriguez \& Chini 1999).  Thus we obtain something like the
Kolmogorov $-5/3$ slope here (for 1-D scans in the projected
distribution of a passive scalar) even though many of the stars in
the optical images may have been triggered along sharp
fronts by nearby high pressures.

\section{Flocculent Spirals inside the Grand Design Galaxy M81}
\label{sect:m81}

Figure \ref{fig:m81floc} shows the galaxy M81 without the 10
lowest Fourier components.  A few bright stars have also been
removed.  The remaining part of the galaxy image lies mostly in
the power-law region of the power spectrum. The star formation
features are patchy, sheared spirals that resemble flocculent
spirals in other galaxies.  They tend to be concentrated in the
density-wave spiral arms (see Fig.\ref{fig:ellipse}) because of
the flow pattern, which moves the gas slower through the arms than
the interarm regions. The flow also has an arm region with reverse
shear, which gives some of the patches larger pitch angles than
the arms (Balbus 1988; Kim \& Ostriker 2001). Nevertheless, the
intrinsic structure of these star-forming patches is hierarchical
and self-similar, as indicated by the power spectrum, so the star
formation process inside of them is more closely related to
turbulence than to spiral waves.  The detailed star formation
processes are therefore the same in flocculent and grand design
galaxies.

\section{Discussion}
\label{sect:sum}

The spiral arms in flocculent galaxies and the star-formation
patches in grand design galaxies appear to be the largest optical features
generated by turbulence with a near-Kolmogorov scaling law. The
similarity between the azimuthal sizes of these regions
and the inverse of the critical Jeans wavenumber suggests they form by
gravitational instabilities in the ambient gas. These
instabilities in the presence of shear compress and twist the gas
into flocculent spiral arms, and they also generate supersonic
turbulence, which cascades to smaller scales,
making a hierarchy of cloud and star-formation structures.  The
reverse process of cloud destruction, which is required to ensure
a steady state, is the role of star formation. The
pressures associated with star formation also contribute
to small scale turbulence.

Hierarchical structures for star formation have not yet been found
in numerical simulations of galaxies or clouds, although this
feature of turbulent gas alone has been simulated by several
groups (MacLow \& Ossenkopf 2000; Klessen 2000; Ossenkopf,
Klessen, \& Heitsch 2001; Ostriker, Stone, \& Gammie 2001).
Hierarchical structure in collapsing gas was simulated by Semelin
\& Combes (2000), Huber \& Pfenniger (2001a, 2002), and Chavanis
(2002). The simulations that are most relevant to what we find
here were by Huber \& Pfenniger (2001a), who got a fractal
mass-size correlation in the spirals generated by gravitational
instabilities in sheared gaseous disks, although their dynamic
range was small.  Wada \& Norman (2001) also got chaotic spirals
from disk instabilities; their effective driving length was
$\sim200$ pc, similar to that observed here, but these authors did
not analyze the correlations in their gas or star formation
structures.

The turbulent spirals in 2-D simulations by Wada, Meurer, \&
Norman (2002) were found to have a 2-D energy power spectrum with
a power-law slope of $\sim-1$ due to inverse cascade on scales
larger than the driving scale of 10 pc, which was from
self-gravity in the cold gas component.
Some of the power law range for structure observed optically here
could be from an inverse cascade too,
but the difference between this process and stretching from shear
is not clear from the 2-D analysis in Wada et al., and the
importance of shear has been minimized in our study by
considering only the azimuthal profiles for the power spectra.
The importance of inverse cascades in galaxies with a
scale height larger than the Jeans length in the cold
component is also not clear.

HST archival images were prepared by the staff at the Space
Telescope Science Institute for NASA under contract NAS5-26555. We
are grateful to J.-C. Cuillandre for the image of M81. Funding for
one of us (S.N.L.) was provided by the Keck Northeast Astronomy
Consortium for summer work at Vassar College. B.G.E is supported
by NSF grant AST-0205097.  We are grateful to the referee for
helpful suggestions.

\begin{deluxetable}{llllllll}
\scriptsize \tablecaption{Galaxies} \label{tab:gal}
\tablewidth{0pt} \tablehead{ \colhead{Galaxy} & \colhead{Passband}
& \colhead{Hubble} & \colhead{Arm} &
\colhead{Distance\tablenotemark{c}} & \colhead{Pixel Size} &
\colhead{Pixel Size} & \colhead{$\Delta R$\tablenotemark{d}}\\
\colhead{} &  \colhead{} &  \colhead{Type\tablenotemark{a}} &
\colhead{Type\tablenotemark{b}} & \colhead{(Mpc)} &
\colhead{arcsec} & \colhead{parsecs}&\colhead{parsecs} }
\startdata
NGC 3031       &  B   & SA(s)ab    &G& 1.4  & 1.0 & 7.1 & 508 \\
NGC 4414       &  R   & SA(rs)c    &F& 9.7  & 0.10 & 4.7 & 225 \\
NGC 5055       &  I,B & SA(rs)bc   &F& 7.2  & 0.10 & 3.5 & 167 \\
NGC 5253       &  V   & Im pec     &F& 3.2  & 0.10 & 1.5 & 49 \\
NGC 7217       &  R   & (R)SA(r)ab &F& 16.0 & 0.10 & 7.7 & 308 \\
NGC 7742       &  I,V & SA(r)b     &F& 22.2 & 0.10 & 10.8 & 343 \\
\enddata
\tablenotetext{a}{From RC3, de Vaucouleurs et al. (1991)}
\tablenotetext{a}{F = flocculent, G = grand design}
\tablenotetext{c}{Distances from Tully (1988)}
\tablenotetext{d}{Separation between azimuthal scans in the radial direction}
\end{deluxetable}

\begin{deluxetable}{lllll}
\scriptsize \tablecaption{Maximum $1/k$ Sizes for Kolmogorov Power
Law} \label{tab:sizes} \tablewidth{0pt} \tablehead{
\colhead{Galaxy} & \colhead{Small Radius} &\colhead{}
&\colhead{Large Radius}&\colhead{}\\
\colhead{}&\colhead{parsecs}&\colhead{R/R$_{25}$}&\colhead{parsecs}&\colhead{R/R$_{25}$}}
\startdata
NGC 3031 (M81) & 500 & 0.1 & 100 & 0.02 \\
NGC 4414       & 300 & 0.06 & 100 & 0.02 \\
NGC 5055       & 1000 & 0.07 & 50 & 0.003\\
NGC 5253       & 60  & 0.03 & -- & -- \\
NGC 7217       & 200 & 0.02 & 60 & 0.007 \\
NGC 7742       & -- & -- & -- & -- \\
\enddata
\end{deluxetable}

\newpage
\begin{figure}
%\plotone{f1.eps} 
\caption{(Left) Power spectra of azimuthal
profiles of HI emission from the Large Magellanic Cloud, using
data from Elmegreen, Kim \& Staveley-Smith (2001).  Each curve is
for a different radius, as indicated. The power spectrum of the
sky is shown as a dashed line for comparison. Lines with slopes of
$-1$, $-5/3$ and $-8/3$ are overlayed. (Right) The same power
spectra are shown again, multiplied by $k^{5/3}$ to flatten the
Kolmogorov part.  Now the sky noise has a rising curve with a
slope of $2/3$ and the $-8/3$ slope in the original power spectrum
has a slope of $-1$, as shown by the short decreasing lines on the
right.  The increase at the far right of these normalized power
spectra is probably noise, since it has the same slope as the sky
noise. } \label{fig:lmc}
\end{figure}

\begin{figure}
%\plotone{f2.eps} 
\caption{Images of the galaxies used for this
study. The NGC numbers and lengths of one kpc are indicated.
All but NGC 3031 are flocculent.  NGC 3031 (orientation is North down)
is from J.C. Cuillandre
of the CFHT; the others are from the HST archives. }
\label{fig:images}
\end{figure}

\begin{figure}
%\plotone{f3.eps} 
\caption{(Bottom) Average azimuthal scans of M81
at a radius of 2 kpc, with a star present on the left and the star
mostly removed on the right.  Each of these scans is an average of
nine pixel-wide azimuthal scans that are separated by one pixel in
radius. (Top) The decreasing curve on each side is the average of
nine power spectra, one for each of the azimuthal scans that goes
into the average shown at the bottom.  The flat curves are the
power spectra multiplied by the $5/3$ power of the spatial
frequency, or wavenumber $k$.}
\label{fig:azim}
\end{figure}

\begin{figure}
%\plotone{f4.eps}
\caption{Negative images of M81 (left) and NGC
5055 are shown with ellipses at the radii of the azimuthal scans
and power spectra. } \label{fig:ellipse}
\end{figure}

\begin{figure}
%\plotone{f5.eps} 
\caption{Normalized power spectra for M81 and NGC
4414 at different radii, increasing toward the top.  The radial
intervals are given in Table 1: 508 pc starting at 508 pc for M81,
and 225 pc starting at 225 pc for NGC 4414. The dashed line has a
slope on this figure of $2/3$, which corresponds to $-1$ in the
original power spectrum, before normalization with the factor
$k^{5/3}$.  The scales at the top of the figure correspond to
inverse wavenumber, $1/k$, and are in parsecs and fractions of the
galaxy radius, R$_{25}$.  The scale at the bottom is normalized
wavenumber, which has $k=1/(2\;{\rm px})$ at the right edge for
each scan (where the normalized wavenumber = 1).  The power
spectra for the outer parts of M81 increase sharply at low $k$, by
two orders of magnitude compared to the extrapolation from higher
$k$.  This excess power is from the spiral density wave. The power
spectra for M81 at intermediate-to-high $k$, which are dominated
by star formation, are about the same as for the flocculent galaxy
NGC 5055.} \label{fig:psm81}
\end{figure}

\begin{figure}
%\plotone{f6.eps} 
\caption{Normalized power spectra for NGC 5055 in
two passbands. The radii are in increments of 167 pc. The inner
regions (lower curves) have a long flat part on this diagram,
suggesting optical features with the same structure as Kolmogorov
turbulence.  A three-segmented line illustrates this flat part, in
comparison to a rising part, probably from noise, at lower
wavenumber, and a falling part, which may be from the resolved
line-of-sight disk thickness, at higher wavenumber.  The rise at
highest wavenumber has the same slope as the noise.}
\label{fig:psn5055}
\end{figure}

\begin{figure}
%\plotone{f7.eps} 
\caption{Normalized power spectra for NGC 5253
and NGC 7217 at radial intervals of 49 pc and 308 pc. The
innermost regions (lower curves) have parts with power spectra
similar to that of Kolmogorov turbulence.  } \label{fig:psn5253}
\end{figure}
\clearpage

\begin{figure}
%\plotone{f8.eps} 
\caption{Normalized power spectra for NGC 7742 in
two passbands. The radii are in increments of 343 pc. There is no
indication of Kolmogorov turbulence in the optical structure of
this galaxy.} \label{fig:psn7742}
\end{figure}

\begin{figure}
%\plotone{f9.eps} 
\caption{Image of M81 (North down)
with the lowest 10 wavenumbers
removed ($k=1,2,...10$).  The bright features are star formation regions and
the dark features are dust or low-intensity disk regions. Several bright
foreground stars have been removed.  The star formation structure in this
image is flocculent and similar to that of NGC 5055, although for M81 it is concentrated
in the regions of the spiral density wave arms.} \label{fig:m81floc}
\end{figure}

\end{document}